\documentclass[]{spie}  %>>> use for US letter paper
\usepackage[]{graphicx}
\usepackage{amssymb}

\title{Nash equilibrium in quantum superpositions} 

\author{Faisal Shah Khan and Simon J. D. Phoenix  \\
Khalifa University of Science, Technology \& Research
PO Box 127788, Abu Dhabi, UAE 
}
\authorinfo{The authors are members of the Quantum Engineering Group at Khalifa University. Email for F. S. Khan: faisal.khan@kustar.ac.ae. Email for S. Phoenix: simon.phoenix@kustar.ac.ae }

\begin{document}
\maketitle

\begin{abstract}  
A working definition of the term ``quantum game'' is developed in an attempt to gain insights into aspects of quantum mechanics via game theory.  
\end{abstract}

%>>>> Include a list of keywords after the abstract 


%%%%%%%%%%%%%%%%%%%%%%%%%%%%%%%%%%%%%%%%%%%%%%%%%%%%%%%%%%%%%
\section{INTRODUCTION}
\label{sec:intro}  % \label{} allows reference to this section

The last decade or so has seen much excitement and controversy regarding the term ``quantum game'' mainly due to the apparent plethora of perspectives that exist which in turn have produced many different working definitions of the term. However, upon close inspection of the existing literature, two main perspectives regarding quantum games can be identified. One, first offered in the original works of Meyer \cite{Meyer} and Eisert, Wilkens, and Lewenstien \cite{EWL}, essentially views a quantum game as the result of higher order randomization in a game via quantum superposition followed by measurement, a generalization of von Nuemann's approach of randomizing in a game via probability distributions with the goal of observing better-paying Nash equilibria. This perspective puts the theory of quantum games in a strictly game-theoretic context as a study of enhanced game-theoretic results via quantum mechanics. The second perspective, also proposed by Meyer \cite{Meyer1}, views quantum games as an aspect of quantum information processing. In particular, Meyer proposes that certain quantum games can be viewed as quantum algorithms. This perspective puts the theory of quantum games strictly in a quantum mechanical context as a study of the control of quantum information processing via game theory. 

Bleiler \cite{Bleiler} argues that in the game-theoretic context, it is more accurate to refer to quantum games as ``quantized games'', a term that follows naturally from the term ``mixed games'' used for games in which randomization via probability distributions is utilized and which more accurately reflects the underlying game-theoretic context. The study of quantized games then explores the possibility of gaining new insights into game-theory via quantum mechanics. 

In the quantum mechanical context where the strict game-theoretic context is secondary, quantum games should perhaps be referred to as ``gamed quanta''. As inelegant as this term is, it more accurately reflects the underlying quantum mechanical context. However, instead of introducing inelegant terminology for the sake of contextual clarity, we adhere to the original term ``quantum games'' to describe the attempt at gaining insights into quantum mechanics via game theory, but insist that any working definition of the term accurately reflect the quantum mechanical context. We propose a working definition in section \ref{sec:Quantum} which, while reminiscent of Meyer's work \cite{Meyer1}, is more general and based on the systematic approach adopted by Telgarsky \cite{Telgarsky} in his development of a working definition of a topological game. To this end, we first giving a brief review of quantized games.  

\section{QUANTIZED GAMEs}\label{sec:Quantized}

Multi-player game theory can informally be described as the mathematical study of conflict and cooperation between various interacting individuals. Call the interaction a {\it game}, the individuals involved {\it players}, and the ability of a player to interact with the other players his {\it pure strategies}. Suppose also that each player has stakes in the game called the {\it payoffs} and that each player is {\it rational}, that is, she will seek to maximize her payoffs in a manner consistent with some preference relation over the payoffs. A play of the game now entails the choice of a pure strategy by each player the result of which is a tuple of pure strategies called a pure strategy profile. Payoff to the players is determined by the particular pure strategy profile employed. In this context, each player will choose a pure strategy that is a best reply to his opponent's choice of pure strategy, thus maximizing his payoff. If every player succeeds in finding such a strategy, then the resulting pure strategy profile is called a {\it Nash equilibrium}. Identification of Nash equilibria is a fundamental goal of multi-player game theory.

In a given game however, Nash equilibria are not necessarily optimal. Worse, they may not even exist. In such cases, Von Neumann calls upon the players to enlarge their strategy sets to include mixed strategies, that is, randomization between their pure strategies via probability distributions. The use of mixed strategies results in probability distributions over the pure strategy profiles and the payoffs are now computed as expected payoffs. When the strategy sets of players are finite, the merit of using mixed strategies arises from Nash's famous theorem which states that equilibrium always exists in terms of mixed strategies. Moreover, it is often the case that such equilibrium is optimal or close to optimal. 

Enlarging the set of strategies available to the players in a game is not merely a time honored-heuristic. It is in fact a mathematically sound procedure in the following sense. A game can be viewed formally in terms of its {\it payoff function} which takes a pure strategy profile to a payoff profile, a tuple of real numbers that assigns to each player the payoff corresponding to the player's particular choice of strategy in the strategy profile. As such, the use of mixed strategies in a game amounts to extending the domain of the payoff function to include probability distributions, resulting in what is often called a ``mixed'' game. This extension is ``proper'', that is, the extended game can always be restricted to the pure strategies to recover the original game. Proper extensions allow for a meaningful comparison between the results generated by the extended game and the original one. From now on, no distinction will be made between a game and its payoff function.

Other extensions are possible. One extension, proposed by Meyer \cite{Meyer} about a decade ago, allows players to utilize pure quantum strategies, that is, sets of qudits together with quantum operations on them. The use of pure quantum strategies results in a higher order randomization between the pure strategy profiles via quantum superpositions which are complex linear combinations followed by measurement, that is, orthogonal projection. Expected payoff is now computed via the probability distribution over pure strategy profiles that results from measurement. Such an extension of a game is known as quantization of the game, and the resulting game itself has been called a ``quantum'' game. Since the fundamental idea behind game quantization is that of forming quantum superposition of pure strategy profiles, Bleiler \cite{Bleiler} has recently proposed that pure quantum strategies be any non-empty set. The area of research that studies quantum games is known as quantum game theory and a major consideration in the subject is the appearance of ``new'' optimal or close to optimal Nash equilibria in terms of quantum strategy profiles.

Unlike extensions to mixed games however, quantizations are not automatically proper. This somewhat subtle fact has in the past led to questions about the relevance of the Nash equilibria that manifest in improperly quantized games to the corresponding classical game. Such issues were recently resolved by Bleiler \cite{Bleiler} via a mathematically formal approach to quantization of games in terms of domain extension. In the language of the Bleiler formalism for ``quantum mixing'', quantizations from which the original game and the mixed game can be recovered upon restriction of the domain are, respectively, {\it proper} and {\it complete} quantizations. Note that a complete quantization is automatically proper. Both proper and complete quantizations make it game theoretically meaningful to speak of ``new'' Nash equilibria in quantized games.

A complete quantization of the popular Hawk-Dove game Prisoner's Dilemma is proposed by Eisert, Wilkens, and Lewenstien \cite{EWL}. These authors show that a new optimal Nash equilibrium appears in the game for quantum strategy profiles consisting of a certain sub-class of quantum strategies. However, when quantum strategy profiles consisting of the most general class of quantum strategies are employed, the only Nash equilibrium that manifests is the sub-optimal one in terms of the players' original pure strategies. However, a further extension of the game to include mixed quantum strategies, that is, probability distributions over the pure quantum strategies of the players, results in a Nash equilibrium in which each player gets a payoff close, but not equal to, the optimal payoff in the game. 

%\begin{figure}
%\centerline{\includegraphics[scale=0.5, bb=5in 1.5in 6.5in 7.7in]{Gquant}}
%\caption{\small{Extension of the game $G$ to $G^{\Theta}$.}}
%\label{Gquant}
%\end{figure}

\section{Quantum Game}\label{sec:Quantum}

Quantized game theory has been shown to hold the potential of offering new insights into game theory via quantum mechanics. The converse question of whether there exist a theory of quantum games where game theory offers new insights into quantum mechanics has been raised frequently since the inception of quantum games, and it seems that Meyer \cite{Meyer1} made an original attempt to answer this question by exploring connections between quantum algorithms and quantum games. In particular, Meyer shows that certain quantum mechanical systems can be viewed as what he calls $PQ$ games, a terminology motivated by the characters of Picard and $Q$ from the television show {\it Star Trek: the next Generation}. He proceeds to show a relationship between certain $PQ$ type quantum games and those quantum algorithms that are modeled as oracle problems such as the Deutsch-Jozsa, Simon and Shor algorithms. Implicit in the details of this work is the point of view that insight into quantum mechanical aspects maybe possible via game theory. In this section, we revisit this point of view by defining a more general working definition of a quantum game following the systematic approach used by Telgarsky \cite{Telgarsky} to develop the notion of topological games.

Formalizing the discussion in section \ref{sec:Quantized} for two player games shows that the mathematical essentials of a multi-player game are the payoff function, call it $G$, that takes the product $S_1 \times S_2$, of the sets of strategies of the players to the set of outcomes $O$ of the game. Most importantly, implicit in this formalism is the notion of a preference relation, call it $P_i$, for each player $i$ over the set of outcomes $O$ (it is when players' preference relations are non-identical that conflict arises in a multi-player games and makes them interesting to study). Note that a multi-player game is defined in terms of set-theoretic objects only. However, if these sets and functions entertain additional mathematical structure, then the game may be given a broader context by utilizing this additional structure in defining the various $P_i$ and the computation of payoffs. When this is the case, the word ``game'' is prefixed with a term that captures the additional mathematical features of the underlying sets and functions. For example, Telgarsky \cite{Telgarsky} defines ``topological'' games in exactly this fashion.  

Now consider a quantum mechanical system described by a unitary map $Q$ that goes from the state space $\mathcal{H} \times \mathcal{H}$ of two spin-half particles (we restrict to two particles for simplicity; what follows extends to the general case) to the joint state space $\mathcal{H} \otimes \mathcal{H}$ of these two particles. Game-theoretic language can be assigned to this quantum system with the domain $\mathcal{H} \times \mathcal{H}$ of $G$ labeled as the product of the strategy sets of the players $A$ and $B$, the function $Q$ as the payoff function, and the range $\mathcal{H} \otimes \mathcal{H}$ of $G$ as the set of outcomes of the players. That is, 
$$
Q: \mathcal{H} \times \mathcal{H} \longmapsto \mathcal{H} \otimes \mathcal{H}
$$
Note however that although the mathematical formalism for $Q$ and $G$ {\it look} similar, what is missing from the formalism for $Q$ as a payoff function is an implicit notion of preference relations over the outcomes. Observing this difference between the formalism of functions $Q$ and $G$ allows us to make the point that simply carrying over part of the language of game theory onto quantum mechanical objects does not capture the essence of what it means to play a game with quantum mechanical objects. 

So how are we to define players' preference relations over quantum superpositions for the payoff function $Q$ so as to produce a working definition of a quantum game that is in line with Telgarsky' approach? We use the notion of distance induced on $\mathcal{H} \otimes \mathcal{H}$ via its inner-product together with measurement. More precisely, fix an orthogonal basis $\mathcal{B}$ of the joint state space $\mathcal{H} \otimes \mathcal{H}$ and define two different orders on its elements. Identify these two different orders as non-identical preference relations of the two players over the elements of $\mathcal{B}$ each ranging from ``least preferred'' to the ``most preferred'' ($M_{\rm player}$) element of $\mathcal{B}$. A player will now prefer an arbitrary quantum superposition $p$ over another $q$ if the distance between $p$ and $M_{\rm player}$ is less than the distance between $M_{\rm player}$ and $q$ because, as per the axioms of quantum mechanics, a measurement of $p$ will result in $M_{\rm player}$ with higher probability versus a measurement of $q$.

With the notion of preference relations over outcomes of $Q$ in place, we next revisit and modify the notion of the players' strategies in the formalism for $Q$ with the goal of putting a working definition of a quantum game in place. We re-define the strategy set of each player to be the set $(\mathcal{H}, U)$ where $U$ is the set of unitary functions between the state space of two spin-half particles and the joint state space of the two particles. That is, each players strategy is now some initial state of his particular spin-half particle together with a unitary operation on the state space of both spin-half particles. With this modified definition of strategies, the notion of a payoff function in the quantum mechanical set up here needs modification as well; the payoff function is no longer the unitary map $Q$ between $\mathcal{H}\times \mathcal{H}$ and $\mathcal{H} \otimes \mathcal{H}$ but rather the function $\mathcal{Q}$
$$
\mathcal{Q}: (\mathcal{H}, U) \times (\mathcal{H},U) \longmapsto \mathcal{H} \otimes \mathcal{H}
$$
defined by
$$
\mathcal{Q} \left(|a \rangle, U_1, |b \rangle, U_2 \right) = U_1U_2(|ab \rangle).
$$
We are now in a position to define a quantum game (for two players): a {\it quantum game} is the quantum mechanical system $\left( (\mathcal{H}, U) \times (\mathcal{H}, U), \mathcal{Q}, \mathcal{H} \otimes \mathcal{H}, \mathcal{B} \right)$.

The question that comes up naturally next is whether a suitable notion of Nash equilibrium might be defined in terms of the strategies of the players in a quantum game. An affirmative answer follows by defining a quantum superposition $E \in \mathcal{H} \otimes \mathcal{H}$ to be an {\it equilibrium} if $E$ satisfies the property of being simultaneously closest to the most preferred element of $\mathcal{B}$ of \underline{each} player in terms of the notion of distance on $\mathcal{H} \otimes \mathcal{H}$. That is, upon measurement, $E$ satisfies both
\begin{equation}\label{conditions} 
|E \cdot M_A|^2 \geq |S \cdot M_A|^2 \hspace{.1in} {\rm{and}} \hspace{.1in} 
|E \cdot M_B|^2 \geq |S \cdot M_B|^2  
\end{equation}
for any other quantum superposition $S \in \mathcal{H} \otimes \mathcal{H}$. The notion of Nash equilibrium is now defined as a pair of strategies $\left((|q_A\rangle,U_A), (|q_B \rangle, U_B)\right) \in (\mathcal{H}, U) \times (\mathcal{H}, U)$, such that 
$$
\mathcal{Q}\left(|q_A\rangle, U_A, |q_B \rangle, U_B\right)=E
$$

\section{Conclusion}

The meaning of the term ``quantum game'' is explored from the perspective of gaining insights into aspects of quantum mechanics via game theory. From this perspective, a working definition of a quantum game for two players is produced. This definition is motivated by the systematic approach of Telgarsky that uses the additional mathematical features of the underlying sets to define topological games. This requires that a preference relation, one per player, be defined on quantum superpositions. Once this is achieved, notions of strategies and Nash equilibrium follow. Our definition of quantum games can potentially be used to describe equilibrium behavior in quantum mechanical systems from a game-theoretic point of view, such as stationary states for an appropriate notion of quantum Markov process.

\end{document}